\documentclass[
    reprint, 
    aps,
    preprintnumbers,
    twocolumn,
    prb,
    superscriptaddress
]{revtex4-2}

\usepackage[utf8]{inputenc}
\usepackage{booktabs}
\usepackage[multiple]{footmisc}
\usepackage{lipsum}
\usepackage{rotating}
\usepackage{perpage}
\usepackage{chronology}
\usepackage{lipsum}

\usepackage{amssymb}
\usepackage{amsbsy}
\usepackage{amsmath}
\usepackage[version=4]{mhchem}
\usepackage{bm}
\usepackage{bbm}
\usepackage{xfrac}
\usepackage{mathtools}

\usepackage{tikz}
\usepackage[T1]{fontenc}
\usepackage{etoolbox}
\usepackage{graphics}
\usepackage{multirow}
\usepackage{url}
\usepackage{hyperref}
\usepackage{placeins}
\usepackage[normalem]{ulem}

\usepackage{color}

\graphicspath{{mainplots/}} 

\newcommand*\vc[1]{\mathbf{#1}}
\newcommand*\tx[1]{\mathrm{#1}}
\newcommand*\wn{cm$^{-1}$}

\begin{document}

\title[]{Exploring Conformational Landscapes Along Anharmonic Low-Frequency Vibrations}

\author{Souvik Mondal}
\author{Michael A. Sauer}
\author{Matthias Heyden}
\email{mheyden1@asu.edu}
\affiliation{School of Molecular Sciences, Arizona State University, Tempe, AZ 85287, U.S.A.}

\date{\today}

\begin{abstract}

We aim to automatize the identification of collective variables to simplify and speed up enhanced sampling simulations of conformational dynamics in biomolecules.
We focus on anharmonic low-frequency vibrations that exhibit fluctuations on timescales faster than conformational transitions but describe a path of least resistance towards structural change.
A key challenge is that harmonic approximations are ill-suited to characterize these vibrations, which are observed at far-infrared frequencies and are easily excited by thermal collisions at room temperature.

Here, we approached this problem with a frequency-selective anharmonic (FRESEAN) mode analysis that does not rely on harmonic approximations and successfully isolates anharmonic low-frequency vibrations from short molecular dynamics simulation trajectories.
We applied FRESEAN mode analysis to simulations of alanine dipeptide, a common test system for enhanced sampling simulation protocols, and compare the performance of isolated low-frequency vibrations to conventional user-defined collective variables (here backbone dihedral angles) in enhanced sampling simulations.

The comparison shows that enhanced sampling along anharmonic low-frequency vibrations not only reproduces known conformational dynamics but can even further improve sampling of slow transitions compared to user-defined collective variables. 
Notably, free energy surfaces spanned by low-frequency anharmonic vibrational modes exhibit lower barriers associated with conformational transitions relative to representations in backbone dihedral space.
We thus conclude that anharmonic low-frequency vibrations provide a promising path for highly effective and fully automated enhanced sampling simulations of conformational dynamics in biomolecules.
\end{abstract}

\maketitle

\section{Introduction}
\label{s:intro}
The structure-function paradigm in structural biology states that the function of a protein is determined by its three-dimensional structure\cite{maynard1970,anfinsen1973,redfern2008}.
However, growing evidence continues to show that proteins are dynamically active biological assemblies with a strong connection between structure and dynamics\cite{henzler2007,boehr2006}.
Many biological functions of proteins such as molecular recognition\cite{frauenfelder1991}, self-assembly\cite{mcmanus2016}, and enzymatic catalysis\cite{benkovic2003}, are strongly associated with 
conformational changes.
The mechanistic characterization of conformational transitions in proteins and a prediction of their impact on biomolecular function remains a significant challenge despite numerous potential applications.
For example, the ability to include dynamics in computational protein design may lead to improved catalytic activities for artificial enzymes\cite{kiss2013computational,mak2014computational}.
Likewise, modulation of conformational dynamics in a protein drug target through ligand binding promises new pathways for the development of therapeutics, e.g., small molecule drugs with allosteric mechanisms\cite{doerr2016,dror2013}.

Several experimental methods can detect the coexistence of protein conformations or transitions between conformations with varying temporal and spatial resolution. Examples include single-molecular fluorescence, NMR, electron spin resonance, small angle x-ray scattering, cryo-electron microscopy, and transient or two-dimensional spectroscopy\cite{kennis2002,boehr2006,mittermaier2006new,cho2010,thielges2012,reppert2016computational,ramos2019site,mazal2019,matsumoto2021,hasanbasri2023}.
Combined, these methods can describe a rich spectrum of dynamical processes in proteins on timescales ranging from picoseconds to seconds\cite{wedemeyer2002,boehr2006}.
Molecular dynamics (MD) simulations can provide microscopic insights into such processes with atomistic resolution\cite{karplus2002}.
However, conformational dynamics associated with protein function often takes place on timescales of microseconds and beyond, which remains a significant challenge for unbiased simulations despite continuing advances in computer hardware and simulation algorithms\cite{dror2012,shaw2021,ayaz2023,greisman2023}.
Therefore, the development of enhanced sampling methods to speed up the sampling of slow dynamic processes in molecular simulations is an active field of research\cite{bernardi2015,mitsutake2001,torrie1977,isralewitz2001,schlitter1994,henin2004,laio2002,sugita1999,hamelberg2004}.

Current enhanced sampling methods can be categorized in two separate groups: 1) unbiased methods that accelerate the crossing of kinetic barriers in general, e.g., replica exchange MD\cite{sugita1999,wang2011replica} and Gaussian accelerated MD\cite{miao2015gaussian,wang2021gaussian}; 2) biased simulation methods that require reaction coordinates as input. Reaction coordinates are collective variables (CVs) that depend on atomic coordinates and are aligned with slow dynamic processes to be observed. Examples of enhanced sampling strategies that rely on CVs are: steered MD\cite{isralewitz2001}, Umbrella sampling\cite{torrie1977}, metadynamics\cite{laio2002}, weighted ensemble methods\cite{zwier2015westpa}, and others\cite{allen2009forward}.

Enhanced sampling methods based on CVs are best equipped to sample the slowest processes in a biomolecular system. 
However, the selection of suitable CVs is critical for the sampling performance and affects barrier heights and other features of free energy surfaces, while at the same time being highly non-trivial.
Various approaches are frequently used to identify suitable CVs based on structural information or dynamics sampled in unbiased simulations. 
Structure-based approaches aim to identify low-frequency vibrations as candidates for CVs associated with large-amplitude motions and conformational transitions. 
CVs describing the vibrations are obtained from harmonic normal mode analysis with a model of the potential energy of the system\cite{go1983,aalten1995}.
Such models can be based on coarse-grained representations of proteins, e.g., in elastic network models of proteins, or feature atomistic detail. 
However, the harmonic approximations implied for this analysis are most applicable at high frequencies and become increasingly invalid with decreasing frequency in biomolecular systems\cite{hayward1995,ma2005,sauer2023}. 
Extracting CVs directly from dynamic simulations can avoid such problems but often faces challenges from sampling limitations. 
Principal component analysis (PCA) identifies CVs associated with the largest variance, e.g., the largest amplitude motion, observed in the simulation\cite{go1983,garcia1992large,amadei1993essential}, while time-lagged independent component analysis (TICA) extracts CVs associated with the slowest observed dynamics\cite{naritomi2013,noe2015kinetic}. 
Both approaches work well for simulation trajectories of sufficient length to observe the dynamic processes of interest.
However, for protein conformational transitions on timescales of multiple microseconds or milliseconds, obtaining such trajectories comes at a considerable computational cost.
If simulation trajectories are too short, the analysis depends on rare events that are not reproducible in independent sets of simulations.
On the shortest timescales, PCA becomes identical to quasi-harmonic normal mode analysis\cite{levy1984quasi,levy1984quasi2}, which implies harmonic approximations that break down for the lowest frequencies\cite{sauer2023}.
More recent approaches utilize machine learning to predict CVs to advance the sampling of the system based on the dynamics sampled in short trajectories and may provide a powerful alternative\cite{ribeiro2018reweighted,wang2019past,wang2020machine}.

Here, we explore a different alternative to identify CVs associated with conformational transitions. 
Our approach is based on the general idea that low-frequency vibrations are likely candidates for reaction coordinates associated with conformational transitions.
However, in contrast to harmonic or quasi-harmonic normal mode analysis, we utilize our recently proposed FREquency-SElective ANharmonic (FRESEAN) mode analysis\cite{sauer2023} that eliminates the need for harmonic approximations\cite{sauer2023}.
FRESEAN mode analysis is based on a time correlation formalism\cite{mcquarrie2000} that combines mass-weighted velocity auto- and cross-correlations. 
Instead of describing all vibrations in a system with a single set of normal modes, FRESEAN mode analysis generates a distinct set of normal modes for each sampled frequency. 
At each frequency, FRESEAN mode analysis simultaneously quantifies the contribution of each mode to the vibrational density of states (VDoS).
This is essential because only a small set of modes accounts for non-zero contributions to the VDoS at any given frequency.
In previous work\cite{sauer2023}, we demonstrated that, in contrast to harmonic and quasi-harmonic normal mode analysis, FRESEAN mode analysis successfully isolates collective degrees of freedom that exhibit fluctuations in a narrow band around a  selected frequency.
Importantly, FRESEAN modes can also isolate collective degrees of freedom associated with zero frequency dynamics that describe diffusive motion or anharmonic low-frequency-vibrations. 

In this study, we report the first enhanced sampling simulations with CVs obtained from FRESEAN mode analysis in combination with well-tempered metadynamics simulations\cite{barducci2008well}. 
We use alanine dipeptide as a test system, which is often used to test enhanced sampling methods\cite{wang2019past,lincoff2016,kumar2023gpu} and exhibits transitions between multiple conformational states.
The associated kinetic barriers are sufficiently high to prevent efficient sampling in standard molecular dynamics simulations on the sub-microsecond timescale at room temperature. 
In contrast to more complex systems, the selection of appropriate CVs 
is straightforward in this case. 
Apart from methyl group rotations, the only dynamic degrees of freedom are the two Ramachandran angles formed by the peptide bonds with the central alanine residue.
The challenge is to identify similarly suitable CVs (or even better-performing CVs with respect to enhanced sampling simulations) in an automated fashion that is transferable to more complex systems.
As shown in the following, FRESEAN mode analysis of short trajectory fragments containing only local fluctuations reliably identify CVs that enable highly efficient sampling of conformational space in biased simulations. 
Notably, we demonstrate that free energy surfaces spanned by these CVs feature significantly lower barriers than the corresponding free energy surfaces obtained in the space spanned by the Ramachandran angles.
\section{Methods}
\label{s:methods}
\subsection{Unbiased Simulations}
Alanine dipeptide consists of a central alanine residue with acetyl and N-methylamide capping groups that form a peptide bond with the N- and C-terminus, respectively.
The peptide was modeled with the AMBER99SB-ILDN forcefield\cite{lindorff2010} and 
solvated with 700 TIP3P\cite{jorgensen1983} water molecules in a 30 \AA~X 30 \AA~X 30 \AA~cubic box. 
All molecular dynamics simulations were performed with the GROMACS 2020.4\cite{abraham2015} software package.
The initial energy system was first minimized with a steepest descent algorithm followed by an equilibration in the isobaric--isothermal (NPT) ensemble at 300 K and 1 bar for 100 ps. 
For equilibration, we used the Berendsen weak-coupling barostat\cite{berendsen1984} with a 2.0 ps time constant to maintain the system pressure, while the temperature was controlled with an ensemble-preserving velocity-rescale algorithm\cite{bussi2007} with a time constant of 0.1 ps. 
This was followed by a production simulation of 10 ns length in the NPT ensemble, in which temperature and pressure were controlled by
a Nos\'{e}-Hoover thermostat\cite{nose1984,hoover1985} with a 1.0 ps time constant and a Parrinello--Rahman barostat\cite{parrinello1981} with a 2.0 ps time constant.
No constraints for intramolecular bonds or angles of the peptide were applied and the simulation time step was set to 0.5 fs to smoothly integrate resulting mid-infrared vibrations. 
Short-ranged electrostatic and Lennard-Jones interactions were treated with a 10 \AA~real--space cutoff with energy and pressure
corrections for dispersion interactions. 
Long--ranged electrostatic interactions were treated with the Particle Mesh Ewald algorithm\cite{darden1993}, using a 1.2 \AA~grid. 
Coordinates and velocities were stored every 4 fs and used for the subsequent analysis. 

\subsection{FRESEAN Mode Analysis of Trajectory Fragments}
To analyze equilibrium fluctuations
of the peptide, we first 
analyzed the time evolution of the peptide backbone dihedrals
in the unbiased simulation.
We then extracted two trajectory fragments of 700 ps (C$_1$) and 400 ps (C$_2$) length that describe fluctuations within two distinct conformational states of the peptide but no transitions between them.
We then performed FRESEAN mode analysis as described in our previous work\cite{sauer2023} independently for both trajectory fragments.
For the FRESEAN mode analysis\cite{sauer2023}, we 
rotated coordinates and velocities of the peptide trajectory into a common frame of reference and 
computed 
a matrix of mass--weighted time cross-correlations between velocities for all degrees of freedom $i$ and $j$ of the peptide
(including
auto-correlations for $i=j$) with a maximum lag time of 2.0~ps:
\begin{equation}
\label{e:corr}
C_{\tilde{\mathrm{v}},ij}(\tau)=\langle \tilde{\mathrm{v}}_\textit{i}(t) \tilde{\mathrm{v}}_\textit{j}(t+\tau)\rangle_\textit{t}
\end{equation}
Here, the angular brackets signify ensemble--averaging over the simulation time, $\tau$ defines the correlation lag time, and 
the mass-weighting of the correlations is included via weighted velocities $\tilde{v}$.
\(\tilde{\mathrm{v}}(t) = \sqrt{m} \cdot \mathrm{v}(t)\).
Each velocity 
time 
correlation 
matrix element is then Fourier transformed to obtain a frequency-dependent velocity correlation matrix, $\vc{C}_{\tilde{\mathrm{v}}}(\omega)$:
\begin{equation}
\label{eqn:vacf-ft}
    C_{\tilde{\mathrm{v}},ij}(\omega) = \frac{1}{2\pi}\int_{-\infty}^{+\infty} \mathrm{e}^{i \omega \tau} C_{\tilde{\mathrm{v}},ij}(\tau) d\tau
\end{equation}
Notably, the trace of $\vc{C}_{\tilde{\mathrm{v}}}(\omega)$, i.e., the sum of Fourier transformed mass-weighted velocity auto correlations, is directly proportional to the vibrational density of states (VDoS)\cite{chakraborty07,heyden10b,mathias11}.
\begin{equation}
\tx{VDoS}(\omega)=\frac{2}{k_B T} \sum_i C_{\tilde{\mathrm{v}},ii}(\omega) 
\label{e:vdos}
\end{equation}
The eigenvectors and eigenvalues of this matrix can be computed for any of the sampled frequencies.
As described in Ref.\citenum{sauer2023}, the eigenvalues obtained for a given frequency $\omega$ describe the contribution of the corresponding eigenvectors to the vibrational density of states (VDoS) at frequency $\omega$.

Here, we focused our analysis on vibrational modes contributing to the VDoS at zero frequency. 
At zero frequency, the FRESEAN modes (eigenvectors) with the six largest eigenvalues describe rigid-body translations and rotations of the alanine dipeptide (see Figure~S1 and Figure~S2 of the Supplementary Information, SI).
The remaining eigenvectors with non-zero eigenvalues describe vibrations with significant contributions to the zero-frequency spectrum, which use as an indication of anharmonic properties.
For both trajectroy fragments (states C$_1$ and C$_2$), we select non-translational and non-rotational FRESEAN modes at zero frequency with the two largest eigenvalues (modes 7 and 8) as CVs for enhanced sampling simulations described below.

\subsection{Metadynamics Simulations}
For each of the conformational states C$_1$ and C$_2$, we performed an independent enhanced sampling simulation using CVs and starting configurations obtained from the corresponding trajectory fragments of 700 and 400~ps length, respectively.
As described above, we selected eigenvectors seven and eight of the cross correlation matrix $\vc{C}_{\tilde{v}}(\omega)$ at zero frequency as CVs in each case, which were combined with the well-tempered (WT) metadynamics simulation protocol\cite{barducci2008well} as implemented in the PLUMED 2.6.6\cite{tribello2014} plugin for GROMACS.
Additionally we also performed a third WT-metadynamics simulation using the two dihedral angles as CVs. 
All metadynamics simulations were 
run for 50 ns in the NPT ensemble using 
the same general simulation parameters as described above for the equilibrium simulations unless otherwise noted.

Gaussian functions were added to the biasing potential every 1.0~ps with an initial height of 1.0~kJ/mol and a width based on fluctuations along the CVs observed in the equilibrium trajectories.
The unitless bias factor was set to 10 for all cases. 
In the WT-metadynamics simulations, we constrained intramolecular bond lengths using the LINCS algorithm\cite{hess1997}, which allowed us to increase the integration time step to 1~fs. 

\section{Results and Discussion}
\label{s:results}
\subsection{Vibrational Analysis of Unbiased Trajectories}
In Figure~1, we illustrate the two dihedral angles phi ($\phi$) and psi ($\psi$) of alanine peptide (panel A) and plotted their time evolution during a 10~ns unbiased simulation (panel B).
It is apparent that rotations around $\psi$ are fast and transitions between favored states at approximately 170~$^\circ$ (conformation C$_1$) and 0~$^\circ$ (conformation C$_2$) are frequently observed on the timescale of the simulation.
In contrast, rotations around $\phi$ are associated with sufficiently high kinetic barriers that impede sampling in unbiased simulations. 
Earlier studies show that spontaneous $\phi$ rotations occur only on the $\mu$s timescale despite the low complexity of the system.\cite{smith1999}

\begin{figure}[ht!]
\begin{center}
\includegraphics[clip=true,width=0.85\linewidth]{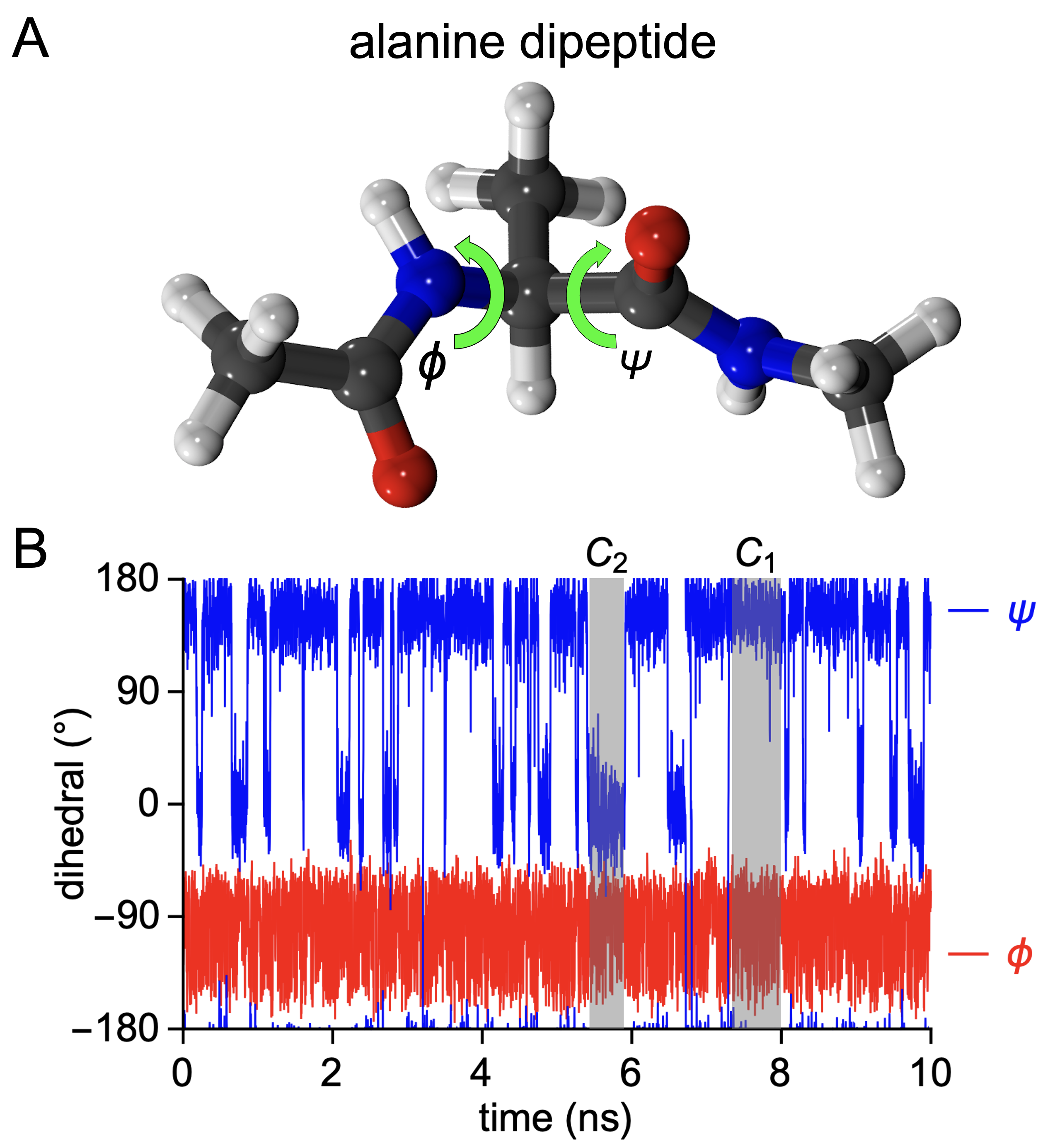}
\caption{a) Structure of alanine dipeptide with backbone dihedral angles $\phi$ and $\psi$, b) Time evolution of $\phi$ and $\psi$ in a 10~ns unbiased simulation with highlighted trajectory fragments selected for two conformational states, C$_1$ and C$_2$, in gray.}
\label{f:trj}
\end{center}
\end{figure}

Selecting the peptide dihedrals as CVs to accelerate the exploration of conformational space is a trivial choice for alanine dipeptide.
However, this system also allows us to test whether low-frequency anharmonic vibrations obtained with the FRESEAN mode analysis formalism can be used to generate suitable CVs without prior information.
The latter is of particular interest for more complex systems such as proteins, for which the identification of CVs associated with conformational transitions is highly non-trivial.

Our goal is to enable the generation of suitable CVs with as little information as possible. 
We thus limit our vibrational analysis of alanine dipeptide to short fragments of the 10~ns trajectory in Fig.~\ref{f:trj}B that exclusively describe local fluctuations within a single conformational state, i.e., either C$_1$ or C$_2$.
We highlighted two selected trajectory fragments in Fig.~\ref{f:trj}B that are 700~ps (C$_1$) and 400~ps (C$_2$) in length.

\begin{figure}[ht!]
\begin{center}
\includegraphics[clip=true,width=0.85\linewidth]{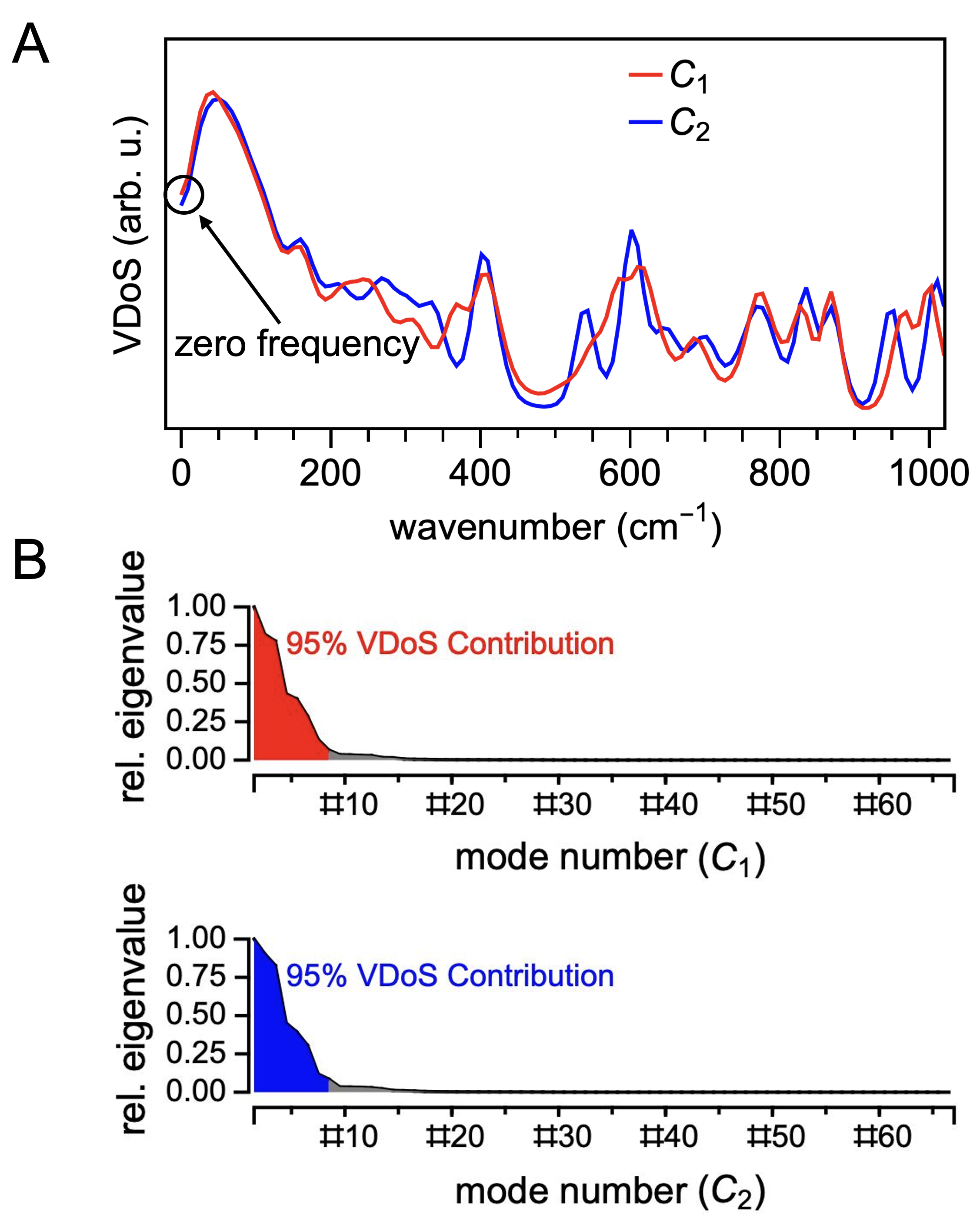}
\caption{A) Vibrational density of states (VDoS) of alanine dipeptide from 0 to 1000 cm$^{-1}$ computed from trajectory fragments C$_1$ and C$_2$. B) Normalized eigenvalues of the frequency-dependent velocity cross-correlation matrix for 0 THz frequency computed for C$_1$ and C$_2$.}
\label{f:vdos}
\end{center}
\end{figure}

We then performed FRESEAN mode analysis independently for these two trajectory fragments. 
First, we plotted the VDoS of the alanine dipeptide for C$_1$ or C$_2$ at far-infrared frequencies in Figure~\ref{f:vdos}A, which we computed from velocity time auto correlation functions as defined in eq.~\ref{e:vdos}.
While the two conformers exhibit slightly distinct vibrations for wavenumbers greater than 200~\wn, they share a broad vibrational band with a peak at $\approx$ 50\wn.

We then compute the frequency-dependent matrix of velocity time auto and cross correlations for the peptide as defined in eq.~\ref{e:corr} to identify collective modes contributing to the VDoS at each frequency.\cite{sauer2023}
Here, we focus on the eigenvectors of the correlation matrix at zero frequency, which isolate collective degrees of freedom associated with translational/rotational diffusion and low-frequency vibrations.\cite{sauer2023}
The corresponding eigenvalues (shown in Figure~\ref{f:vdos}B) quantify the contribution of each eigenvector (FRESEAN mode) to the zero frequency VDoS.
The eigenvalues decrease rapidly with increasing eigenvector index. 
As a consequence, 95\% of the zero frequency VDoS in Figure~\ref{f:vdos}A are associated with fluctuations along the first 8 FRESEAN modes at zero frequency as shown in Figure~\ref{f:vdos}B.

\begin{figure}[ht!]
\begin{center}
\includegraphics[clip=true,width=1.0\linewidth]{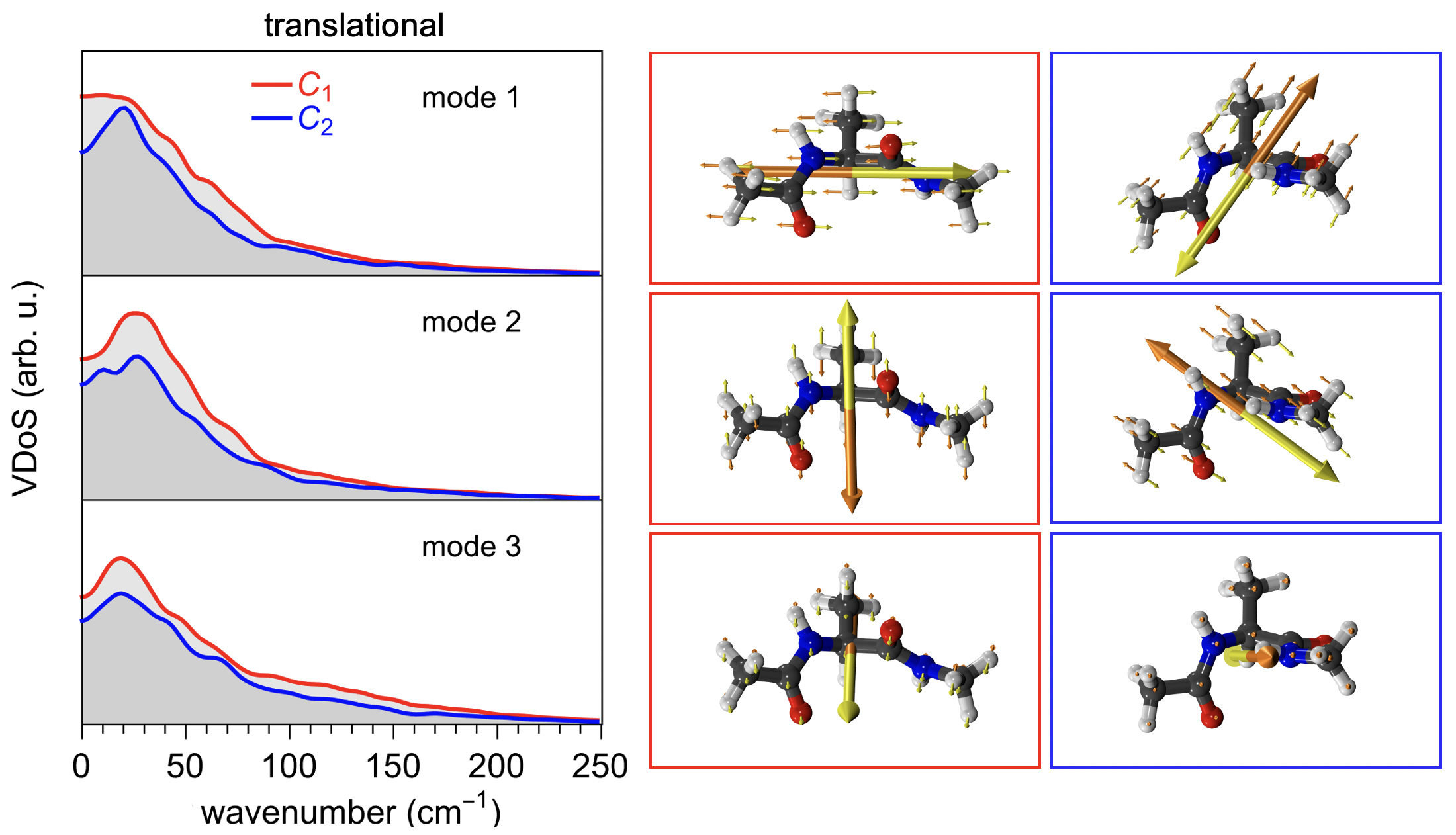}
\caption{
A: 1D-VDoS for projections of trajectory fragments C$_1$ (red) and C$_2$ (blue) on FRESEAN modes 1, 2 and 3 (translations) obtained at zero frequency. B: Visualization of the FRESEAN modes 1, 2 and 3 as displacement vectors relative to the average structure observed in trajectory fragments C$_1$ (left panels, red frame) and C$_2$ (right panels, blue frame). A large bidirectional arrow indicates the corresponding center of mass motion.
}
\label{f:tr}
\end{center}
\end{figure}

\begin{figure}[ht!]
\begin{center}
\includegraphics[clip=true,width=1.0\linewidth]{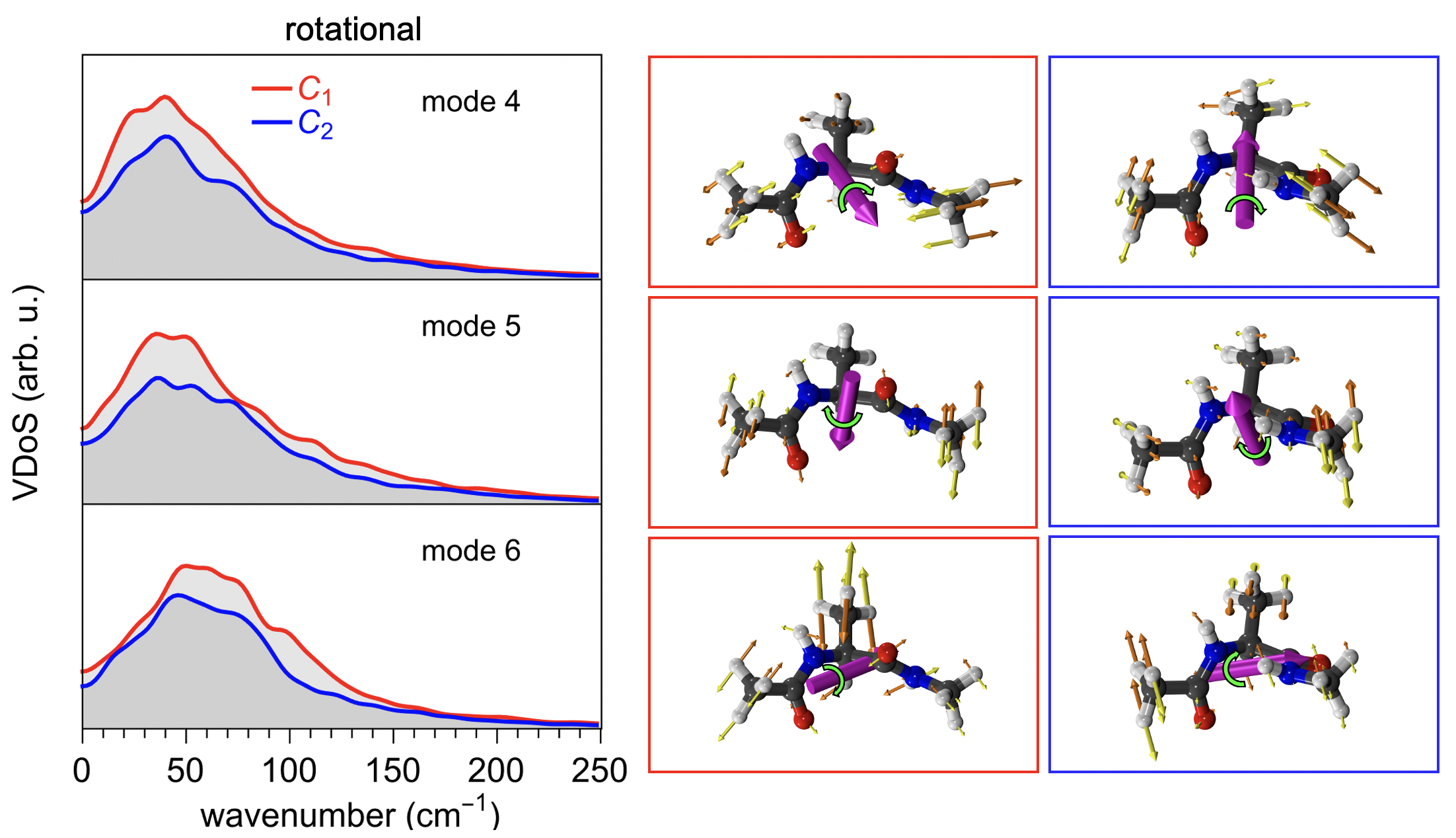}
\caption{
A: 1D-VDoS for projections of trajectory fragments C$_1$ (red) and C$_2$ (blue) on FRESEAN modes 4, 5 and 6 (rotations) obtained at zero frequency. B: Visualization of the FRESEAN modes 4, 5 and 6 as displacement vectors relative to the average structure observed in trajectory fragments C$_1$ (left panels, red frame) and C$_2$ (right panels, blue frame). A large arrow indicates the corresponding axis of rotation.
}
\label{f:rot}
\end{center}
\end{figure}

\begin{figure}[ht!]
\begin{center}
\includegraphics[clip=true,width=0.75\linewidth]
{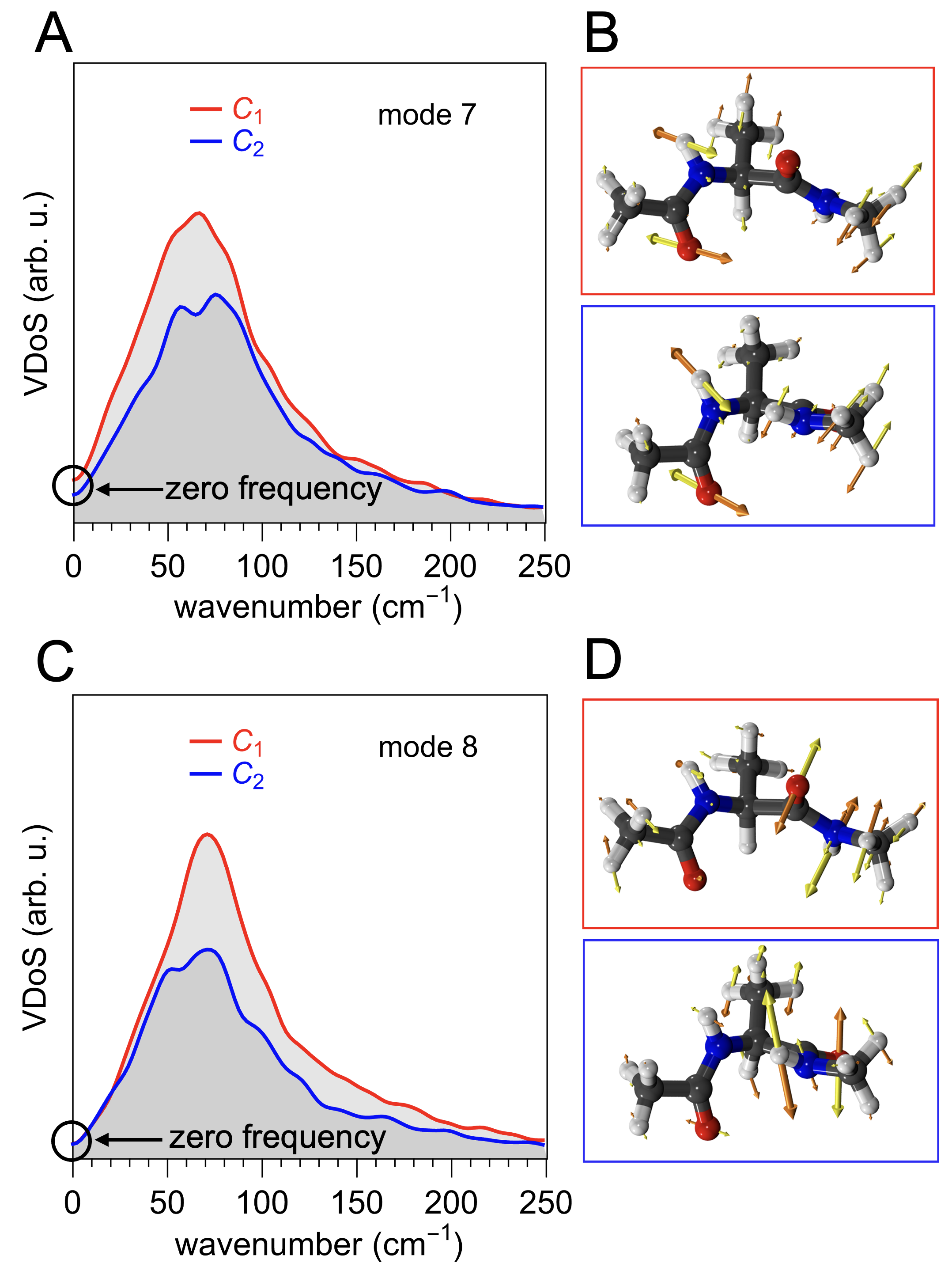}
\caption{A,C: 1D-VDoS for projections of trajectory fragments C$_1$ (red) and C$_2$ (blue) on FRESEAN modes 7 (A) and 8 (C) obtained at zero frequency. The zero frequency contribution to the total VDoS of each mode is highlighted (black circle). B,D: Visualization of the FRESEAN modes 7 (B) and 8 (D) as displacement vectors relative to the average structure observed in trajectory fragments C$_1$ (top, red frame) and C$_2$ (bottom, blue frame).}
\label{f:1d}
\end{center}
\end{figure}

The first 6 modes describe rigid body translations and rotations of the peptide as shown in Figures~\ref{f:tr} and \ref{f:rot}.
Modes 7 and 8 describe low-frequency vibrations with peak frequencies between 50 and 100~\wn and non-zero intensities at zero frequency as shown in Figure~\ref{f:1d}. 
In panels A and C, we plot the isolated 1D-VDoS for fluctuations along modes 7 and 8, respectively, as observed in both analyzed trajectory fragments for conformational states C$_1$ and C$_2$.
The 1D-VDoS are calculated from time auto correlations of projections of weighted velocities ($\tilde{\mathrm{v}}=\sqrt(m) \mathrm{v}$) of the peptide atoms on a given FRESEAN mode $\vc{q}$ with components $q_i$.
\begin{eqnarray}
    \dot{q}(t) &=& \sum_i q_i \tilde{\mathrm{v}}_i(t) \\
    \mathrm{VDoS}_{q}\left(\omega\right) & = & \frac{2}{k_B T} \frac{1}{2 \pi} \int_{-\infty}^\infty \mathrm{e}^{i \omega \tau} \, \langle \dot{q}(t) \dot{q}(t+\tau)\rangle_t \, d\tau
\end{eqnarray}
As for the FRESEAN mode analysis, weighted velocities $\tilde{\mathrm{v}}$ are defined in a reference frame that follows the rotations of the peptide during the simulation. 
We note that apart from the broad main peak, the 1D-VDoS for all four vibrational modes (modes 7 and 8 for C$_1$ and C$_2$) are free from contributions of high-frequency vibrations of the peptide at frequencies outside the range shown in Figure~\ref{f:1d}.
As discussed in our previous work and in contrast to harmonic or quasi-harmonic mode analysis, FRESEAN mode analysis can be used to successfully isolate low-frequency vibrations of molecular systems.\cite{sauer2023}

The broad line shape of the peaks in the 1D-VDoS and the intensity at zero frequency (proportional to the eigenvalue of
$\vc{C}_{\tilde{\mathrm{v}}}\left(0\right)$) 
can be interpreted as indicators of an anharmonic vibration.
We visualized the displacement vectors associated with each vibrational mode in panels B and D of Figure~\ref{f:1d}, which allow for a qualitative characterization of the nature of the vibrations.
For mode 7, in both the C$_1$ and C$_2$ conformations, the dominant feature of the displacement vectors are opposing motions of the carbonyl oxygen and amide hydrogen in the N-terminal peptide bond, which implies a rotation of the $\phi$ angle (see Figure~\ref{f:trj}A). 
Likewise, displacement vectors for mode 8 in both conformations describe opposing motions of the carbonyl oxygen and amide hydrogen in the C-terminal peptide bond as expected for a rotation of the $\psi$ angle.
While each vibrational mode describes a collective motion with contributions from all peptide atoms, both dihedral angles of the peptide clearly have major contributions.
The latter is significant because this information did not enter the analysis and we explicitly analyzed trajectory fragments that did not exhibit large amplitude fluctuations of the dihedrals. 
Hence, the FRESEAN mode analysis identified anharmonic vibrational modes associated with both dihedral angles without prior information purely from local fluctuations.
Notably, this was repeated independently for local fluctuations in conformation C$_1$ and conformation C$_2$.

\subsection{Enhanced Sampling Simulations along Anharmonic Vibrations}
We then tested whether the anharmonic vibrations identified with the FRESEAN mode analysis are suitable CVs in enhanced sampling simulations. 
Specifically, we tested whether CVs based on these vibrations enable efficient sampling of the conformational space of alanine dipeptide.
For this purpose, we performed a set of three independent well-tempered WT-metadynamics simulations as described in the Methods section.
Two simulations started from conformations C$_1$ and C$_2$, respectively, and used the corresponding zero frequency FRESEAN modes 7 and 8 as CVs. 
The third simulation was performed as a reference and used the dihedral angles $\phi$ and $\psi$ as CVs.

\begin{figure}[ht!]
\begin{center}
\includegraphics[clip=true,width=0.85\linewidth]{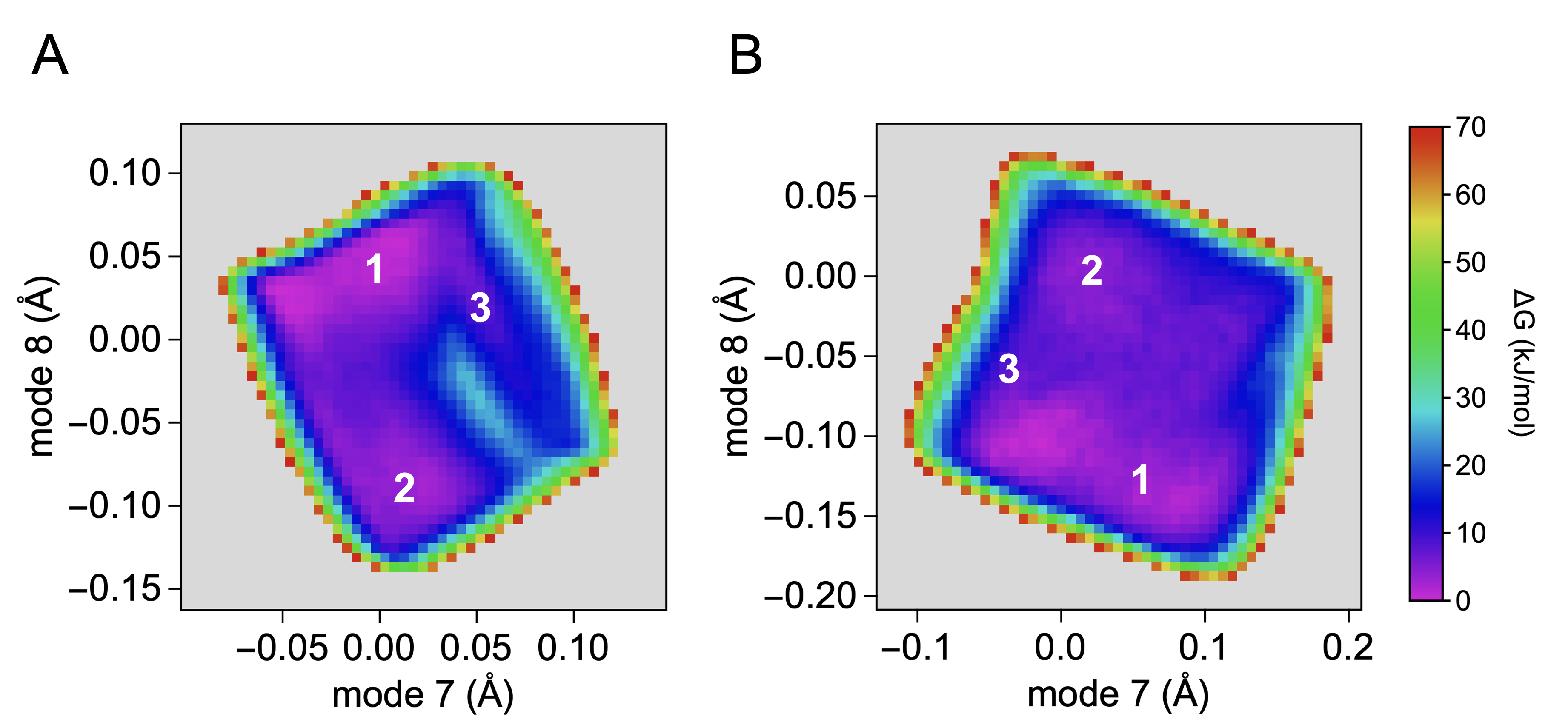}
\caption{Free-energy surfaces obtained from metadynamics simulations with biasing potentials defined by CVs based on FRESEAN modes. 
FRESEAN modes and a reference structure from conformation C$_1$ are used in panel A and from conformation C$_2$ in panel B.}
\label{f:fes1}
\end{center}
\end{figure}

\begin{figure}[ht!]
\begin{center}
\includegraphics[clip=true,width=1.0\linewidth]{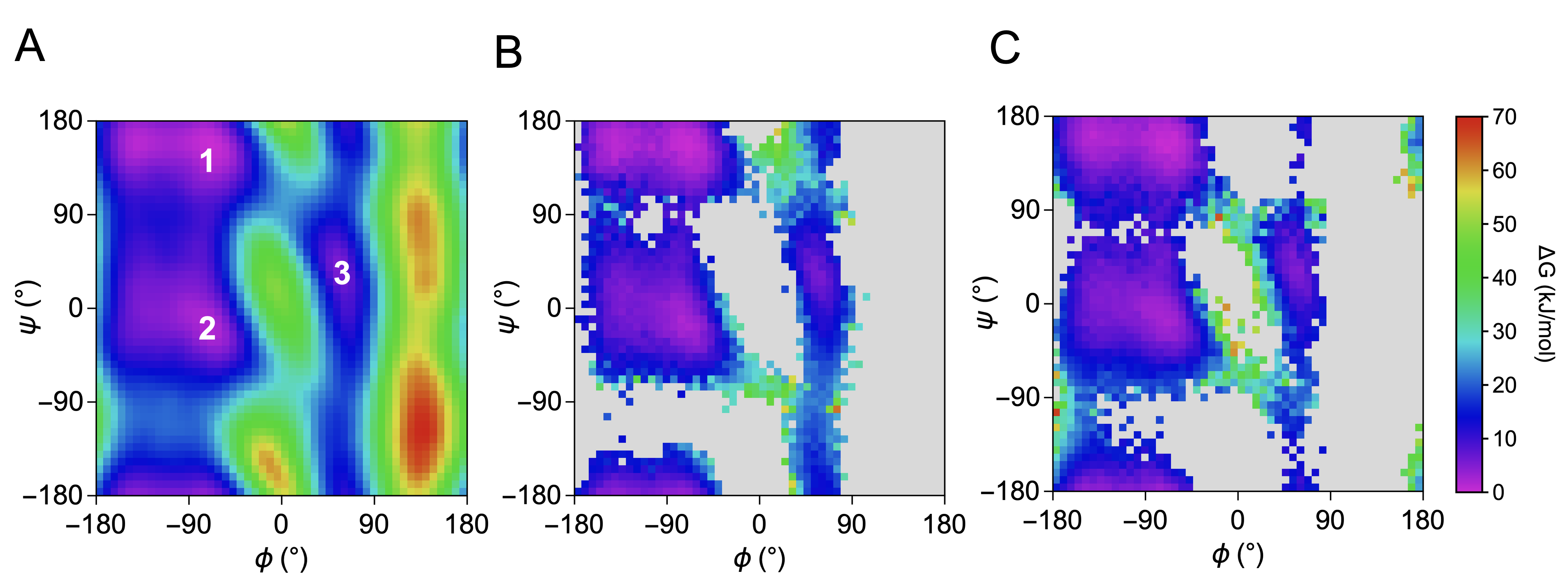}
\caption{Free energy surfaces in dihedral angle space: A) metadynamics simulation with biasing potential defined as a function of the dihedral angles (reference). B) and C) free energy surfaces obtained from unbiased metadynamics trajectories obtained with biasing potentials in FRESEAN mode space.}
\label{f:fes2}
\end{center}
\end{figure}

Free energy surfaces obtained from the first two simulations, which use the FRESEAN modes of conformations C$_1$ and C$_2$ to define the biasing potential, are displayed in Figure~\ref{f:fes1}. 
Differences between both free energy surfaces are expected, because distinct sets vibrational modes obtained from distinct trajectory fragments corresponding to conformations C$_1$ and C$_2$, as well as distinct reference structures, are used to define the projection of atomic coordinates into the FRESEAN mode space.
In Figure~\ref{f:fes2}A, we plotted the free energy surface obtained from the third simulation as a function of the dihedral angles $\phi$ and $\psi$ used as CVs to define the biasing potential in that case.
To allow for direct comparisons between all three simulations, we further plotted free energy surfaces in dihedral angle space in panels B and C of Figure~\ref{f:fes2}, which we obtained after unbiasing the metadynamics simulations obtained with FRESEAN modes as CVs.
For this purpose, we computed the thermodynamic weight $w_k$ of each trajectory time frame ($t_k$) based on its free energy in vibrational mode space (defined by a projection of displacements relative to a reference structure on FRESEAN modes 7 and 8, denoted in the following as $q_7$ and $q_8$).\cite{branduardi2012}
\begin{equation}
w_k = \tx{e}^{-\frac{\Delta G\left(q_7(t_k),q_8(t_k)\right)}{k_B T}}
    \label{e:w}
\end{equation}
Using these weights, we computed the unbiased probability distributions of the dihedral angles, $p\left(\phi,\psi\right)$, and the corresponding free energy surfaces.
\begin{equation}
\Delta G\left(\phi,\psi\right) = - k_B T \, \mathrm{ln} \left[ p\left(\phi,\psi\right) \right]
\label{e:g}
\end{equation}

A notable characteristic of the free energy surfaces in Figure~\ref{f:fes1} is the small magnitude of free energy barriers between distinct minima in the space described by the anharmonic vibrations of conformations C$_1$ and C$_2$.
Several free energy minima are observed that are connected to each other with barriers lower than 10~kJ/mol.
A single barrier of approximately 25~kJ/mol is observed in Figure~\ref{f:fes1}A (based on conformations C$_1$ and its vibrations) for direct transitions between two states indicated as 2 and 3 (more details on this assignment in the following). 
However, no equivalent barrier is seen for indirect transitions passing through state 1 or in the free energy surfaces shown in  Figure~\ref{f:fes1}B (based on conformations C$_2$ and its vibrations).

The latter is especially striking in comparison to the free energy surface obtained in our reference simulation with the backbone dihedral angles $\phi$ and $\psi$ as CVs (Figure~\ref{f:fes2}A).
The latter reproduces multiple previous results in the literature\cite{wang2019past,lincoff2016,kumar2023gpu} and describes free energy barriers below 10~kJ/mol for transitions between states indicated as 1 and 2 (equivalent to conformations C$_1$ and C$_2$ observed in the unbiased simulation trajectory in Figure~\ref{f:trj}) via rotations of $\psi$. 
However, the lowest free energy pathway that connects states 1 and 2 with state 3 in dihedral angle space involves a rotation of $\phi$ with a free energy barrier of 25~kJ/mol.

The simplest explanation for the discrepancy between the observed free energy barriers in Figures~\ref{f:fes1} and \ref{f:fes2}A would be that both metadynamics simulations with CVs based on FRESEAN modes did not explore state 3. 
However, the free energy surfaces obtained after recasting the corresponding metadynamics simulations from FRESEAN mode space into dihedral angle space reproduce the free energy minima associated with each state observed in Figure~\ref{f:fes2}A. 
This includes the shape of each minimum as well as the relative free energy difference between the minima.
Even more noticeable is that the simulations with FRESEAN modes as CVs almost exclusively sample low-free energy regions in dihedral space and avoid high free energy states almost entirely.
This demonstrates that the space spanned by anharmonic low-frequency vibrations identified with the FRESEAN mode analysis describes pathways for conformational changes along which the system experiences minimal restraining forces.

Notably, the absence of high free energy barriers in the space of anharmonic low-frequency vibrations does not contradict the observation of slow kinetics for transitions into state 3 from either state 1 or 2. 
The dynamics of a system on a low-dimensional free energy space can be described within the framework of the generalized Langevin equation, which contains dynamic parameters (diffusion coefficients) that impact the transition rate and are not characterized here.

\begin{figure}[ht!]
\begin{center}
\includegraphics[clip=true,width=1.0\linewidth]{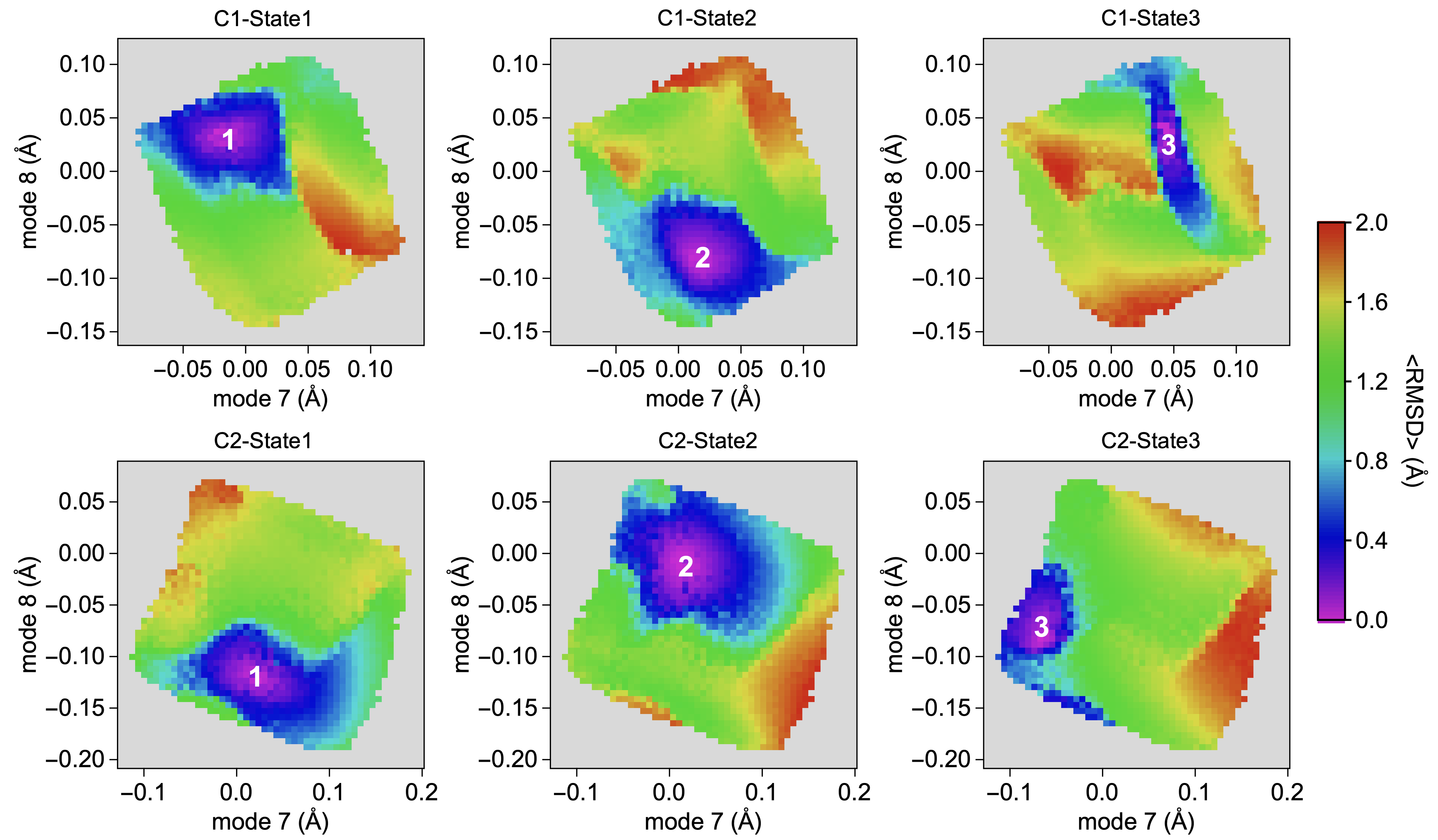}
\caption{
Mapping of conformations sampled in dihedral angle space into FRESEAN modes space: average RMSDs relative to states 1, 2 and 3 indicated in Figure~\ref{f:fes2}A as a function of FRESEAN modes 7 and 8 obtained from conformations C$_1$ (A) and C$_2$ (B).
}
\label{f:rmsd}
\end{center}
\end{figure}

To identify how free energy minima in dihedral angle space are mapped into the FRESEAN mode space, we determined the average structure of configurations corresponding to states 1, 2 and 3 in Figure~\ref{f:fes2}A. 
We then computed the root mean-square displacements (RMSDs) relative to these three structures for all structures observed in the metadynamics simulations in FRESEAN mode space based on C$_1$ and C$_2$, respectively. 
We averaged the RMSDs for all time frames with identical projections in the corresponding FRESEAN mode space and the results are shown in Figure~\ref{f:rmsd}.
The result allows a direct mapping of the free energy minima observed in dihedral angle space in Figure~\ref{f:fes2}A onto distinct regions of the free energy surfaces observed in the FRESEAN mode space in Figure~\ref{f:fes1}A and B.
The latter was used to label the conformational states 1, 2 and 3 from Figure~\ref{f:fes2}A in panels A and B of Figure~\ref{f:fes1}.

\section{Conclusion}
\label{s:con}
The idea to use low-frequency vibrations to guide molecules through conformational space is as old as molecular simulations.\cite{levy1984quasimodes,levitt1985protein,zheng2005normal} 
However, the latter implies a highly anharmonic potential energy surface along such vibrations and a limiting factor for applications has so far been  the lack of theoretical and computational tools to identify vibrations in a regime where harmonic approximations fail.
This has changed with the recent development of the FRESEAN mode analysis, which does not rely on harmonic approximations and identifies vibrational modes contributing to the spectrum at any given frequency.\cite{sauer2023}
The underlying time correlation formalism remains valid at near-zero as well as zero frequencies where it isolates displacement modes that describe translational and rotational diffusion as well as anharmonic low-frequency vibrations.

Notably, anharmonic low-frequency vibrations associated with conformational transitions can be identified from local fluctuations on short timescales that do not describe any conformational transition events. 
Here, we demonstrate this for alanine dipeptide for which we identify low-frequency vibrations associated with conformational transitions on nanosecond and microsecond timescales from simulation fragments of 100s of ps length.
We were further able to repeat this for trajectory fragments describing local fluctuations in two distinct conformational states.
In both cases, we readily identified vibrations describing collective motions that result in dihedral angle rotations to explore the thermally accessible conformational space. 
Enhanced sampling simulations that utilize these vibrations as CVs to define a biasing potential not only speed up conformational sampling. 
These simulations result in free energy surfaces in the space defined by low-frequency vibrational modes that are essentially free from significant free energy barriers.
At the same time, these simulations sample the same conformational space that is also explored when dihedral angles are used directly as CVs to define the biasing potential.
However, in the latter case, free energy surfaces in dihedral angle space include not only substantially higher free energy barriers but also several high free energy states that are not accessible at room temperature and therefore thermodynamically not relevant.

Despite the simplicity of the test system used here, our results show promise for applications in complex biomolecular systems with unknown conformational dynamics. 
Specifically, the ability to identify low-frequency anharmonic vibrations associated with local fluctuations on short timescales could substantially improve our ability to sample functional dynamics in enzymes or to identify allosteric mechanisms in drug targets.
The latter is particularly important with the current drastic increase in structural information for proteins due to advances of structure prediction algorithms\cite{senior2020improved,jumper2021highly,anishchenko2021protein} as well cryo-electron microscopy techniques\cite{yip2020atomic}.

\begin{acknowledgements}
This work is supported by the National Science Foundation (CHE-2154834) and the National Institute of General Medical Sciences (1R01GM148622-01). The authors acknowledge Research Computing at Arizona State University for providing high performance computing resources that have contributed to the research results reported within this work.
\end{acknowledgements}

%

\end{document}